\documentclass[dvips]{article}
\usepackage{latexsym}
\usepackage{amsfonts}
\usepackage{amsmath}
 \usepackage{acronym}
\usepackage{graphics}
\newcommand{\be}{\begin{equation}}
\newcommand{\ee}{\end{equation}}
\newcommand{\bea}{\begin{eqnarray}}
\newcommand{\eea}{\end{eqnarray}}

\makeatletter
\@addtoreset{equation}{section}

 \makeatother
\begin{document}
\normalsize
\title{BPS ansatzes as electric form-factors.}
\author
{\bf L.~D.~Lantsman.\\
  18109, Rostock, Germany; Mecklenburger Allee, 7\\ llantsman@freenet.de\\ Tel. (049)-0381-7990724.}
 \maketitle
 \begin{abstract} We argue that BPS ansatzes, entering manifestly vacuum BPS monopole solutions to equations of motion in the (Minkowskian) non-Abelian Higgs model play the role of some electric form-factors and  that this implies (soft) violating the CP-invariance of the mentioned  model, similar to taking place in the Euclidian Yang-Mills (YM) theory with instantons, generating the $\theta$-term  in the appropriate effective Hamiltonian. 
 \end{abstract}
\newpage 
\tableofcontents
\newpage
\section{Introduction.}
The (without of quarks) non-Abelian YM theory involving vacuum BPS monopole solutions in their Higgs and gauge sectors (as a result of the spontaneous breakdown the initial $SU(2)$ gauge symmetry group to its $U(1)$ subgroup) occupy a special position among another such theories with monopoles. This is associated with manifest superfluid properties of the former model.

\medskip To  elucidate this our assertion, let us at first write down explicitly the action functional for the (Minkowskian) Yang-Mills-Higgs (YMH) model. It can be represented as \cite{Cheng, Ryder, LP1,LP2,rem1}
  \be \label{YM L}
S=-\frac {1}{4 g^2} \int d^4x F_{\mu \nu}^b F_b^{\mu \nu }+ \frac {1}{2} \int d^4x (D_\mu\phi,D^\mu\phi
) -\frac {\lambda}{4} \int d^4x \left[(\phi^b)^2- \frac{m^2}{\lambda}\right]^2,
\ee
with 
$$D_\mu\phi=\partial^\mu\phi+g[A^{\mu },\phi]$$ 
being the covariant derivative an $g$ being the YM coupling constant.\par
The action functional (\ref{YM L}) results the equations of motion \cite{Cheng}
\be \label{eom1}
(D_\nu F^{\mu \nu})_a=-g \epsilon_{abc}\phi^b  (D_\mu\phi)^c,  
\ee
\be \label{eom2}
(D^\mu D_\mu \phi)_a =-\lambda \phi_a ({\vec \phi }\cdot{\vec \phi }-a^2);\quad a^2=m^2/\lambda. 
\ee
\medskip It turns out that going over to the limit 
\be 
\label{lim} 
\lambda\to 0,~~~~~~m\to 0:~~~~~~~~~~ ~~~~~\frac{1}{\epsilon}\equiv\frac{gm}{\sqrt{\lambda}}=ga\not =0; 
\ee
in Eq. (\ref{YM L}) just induces the (topologically degenerated) vacuum BPS monopole solutions in the Higgs and YM sectors of the model (\ref{YM L}).

Historically, the idea to go over to the limit (\ref{lim}) in the YMH model is originated from the works \cite{BPS}, and from this time it was refer to as the {\it Bogomol'nyi-Prasad-Sommerfeld} (BPS) limit.  Later, in the papers \cite{LP1,LP2} the BPS limit was rearranged to the look (\ref{lim}), implicating the YM  coupling constant $g$. 

It is remarkable that the ratio $a=m/\sqrt{\lambda}$ (having the mass dimension) can take arbitrary values  in the limit (\ref{lim}) and that the variable $\epsilon \to 0$ of the length dimension is introduced therein. We shall make sure soon that $\epsilon$ plays the role of the size parameter characterized the core of a BPS monopole.

\medskip Vacuum BPS monopole solutions can be derived in the limit (\ref{lim}) at evaluating the lowest bound of the
energy for the given YMH configuration (often referred to as the {\it Bogomol'nyi bound} in the modern literature) \cite{LP1,LP2}: 
\be 
\label{Emin}
E_{\rm min}= 4\pi {\bf m }\frac {a}{g}
\ee 
(where $\bf m$ denotes  the magnetic charge).

As a result, one arrives at the so-called {\it Bogomol'nyi equation} \cite{LP1,LP2,Gold}
\be
\label{Bog} 
{\bf B}(\Phi) =\pm D \Phi
\ee 
relating the vacuum ``magnetic'' field ${\bf B}$ to the vacuum Higgs  configuration in the shape of a BPS monopole.  The presence of two opposite signs in the Bogomol'nyi equation (\ref{Bog}) corresponds to two opposite
signs of magnetic charges in  nature.

\medskip The explicit way deriving the Bogomol'nyi equation (\ref{Bog}) and evaluating the Bogomol'nyi bound (\ref{Emin}) was stated, for instance, in the monograph  \cite{Al.S.} (see ibid \S $\Phi$11). In particular,  in the monograph  \cite{Al.S.} vacuum  BPS monopole solutions to the Bogomol'nyi equation (\ref{Bog}), arising in the Higgs and gauge sectors of the YMH model \cite{BPS},   were written down. In the series of papers \cite{LP1,LP2} these solutions were  reproduced with the only modification that the effective Higgs mass $a=m/\sqrt{\lambda}$ (utilized in Ref. \cite{Al.S.}) was replaced with the parameter $\epsilon^{-1}$:
\be 
\label{sc monopol}
 \Phi^a_{(0)} (t,{\bf x})    =  \frac{ x^a}{gr} f_0^{BPS}(r)~,~~~~~~~~~~~~~~~~~~~~~~~~~~~~~  
f_0^{BPS}(r)=\left[  \frac{1}{\epsilon\tanh(r/\epsilon)}-\frac{1}{r}\right],  \ee 
\be
 \label{YM monopol} 
 A^a_{i(0)}(t,{\bf x})\equiv\Phi^{aBPS}_i({\bf x}) =\epsilon_{iak}\frac{x^k}{gr^2}f^{BPS}_{1}(r),~~~~~~~~   f^{BPS}_{1}(r)= \left[1 -  \frac{r}{\epsilon \sinh(r/\epsilon)}\right].
\ee 
Indeed, the BPS monopoles (\ref{sc monopol}), (\ref{YM monopol}) are topologically trivial fields \footnote{The topological degeneration of vacuum BPS monopole data (\ref{sc monopol}), (\ref{YM monopol}) can be carry out by means of ``large'' gauge transformations (in the terminology \cite{Fadd2}) proposed in the work \cite{Pervush2}:
$$ { \Phi_i} ^{a(n)}:= v^{(n)}({\bf x})[{ \Phi_i} ^{a(0)}+\partial _i]v^{(n)}({\bf x})^{-1},\quad v^{(n)}({\bf x})=
\exp [n\hat \Phi _0({\bf x})]; $$
$$ \Phi_{(n)a}= v^{(n)}({\bf x}) \Phi_{(0)a} v^{(n)}({\bf x})^{-1}; \quad n\in {\bf Z} $$
(latter Eq. was derived in the paper \cite{Arsen}).

Here
$$ {\hat \Phi}_0(r)= -i\pi \frac {\tau ^a x_a}{r}f_{01}^{BPS}(r); \quad 
f_{01}^{BPS}(r)=[\frac{1}{\tanh (r/\epsilon)}-\frac{\epsilon}{r} ]=f_{1}^{BPS}(r)/\epsilon;  $$
with $\tau ^a$ ($a=1,2,3$) being the Pauli matrices. The exponential multipliers $v^{(n)}({\bf x})$ were referred to as {\it Gribov topological multipliers} in Ref.  \cite{Pervush2} while the value ${\hat \Phi}_0(r)$ as the {\it Gribov phase}. 

It can be argued that specified in this way topologically degenerated YM vacuum BPS monopole data ${ \Phi_i} ^{a(n)}$ satisfy the Coulomb gauge $D^i { \Phi_i} ^{a(n)}=0$. And moreover, YM vacuum BPS monopole data ${ \Phi_i} ^{a(n)}$ also turn out to be gauge invariant, i.e. {\it physical}, functionals of YM fields. In particular, this is correctly for topologically trivial YM BPS monopoles  (\ref{YM monopol}) \cite{LP1,LP2}, and  this   indicates transparently the purely physical nature of these monopoles. Also  topologically trivial Higgs BPS monopole modes (\ref{sc monopol}) and their Gribov  topological copies $\Phi^{(n)a}$ \cite{Arsen} prove to be manifestly gauge invariant,  side by side with YM BPS monopoles. 

\medskip Mention  that the topologically degenerated YM vacuum BPS monopoles ${ \Phi_i} ^{a(n)}$  are patterns of {\it topological Dirac variables},  gauge invariant and transverse (in the sense satisfying the Lorentz covariant Coulomb gauge $D^i { \Phi_i} ^{a(n)}=0$).  

The important point here that topological Dirac variables $\Phi^{(n)a}$ are got indeed as solutions to the YM Gauss law constraint \cite{Pervush2}
$$ \frac {\delta W}{\delta A^a_0}=0 \Longleftrightarrow [D^2(A)]^{ac}A_{0c}= D^{ac}_i(A)\partial_0 A_{c}^i. $$
As  to Higgs BPS monopole modes $\Phi^{(n)a}$, the appropriate current $\rho^H\sim ig \Phi D \Phi$  decouples from the YM Gauss law constraint in the first order of the perturbation theory by the YM coupling constant $g$. 

 The detailed analysis of  topological Dirac variables (including the answer why Higgs BPS monopole modes $\Phi^{(n)a}$ disappear from the YM Gauss law constraint in the first order of the perturbation theory) was performed in the works \cite{LP1,LP2,Pervush2, David2,David3}, and we recommend these to our readers for studying the matter. 
}. 

\medskip Now let us discuss the behaviour of the {\it BPS anzatses} $f_0^{BPS}(r)$ and $f^{BPS}_{1}(r)$ at the origin of coordinates and at the spatial infinity. Direct checking shows that \cite{David3}
\be \label{bcf0}
f_0^{BPS}(0)=0,~~~~~~~~~~~~~~
f_0^{BPS}(\infty)=1;
\ee
\be \label{bcf1}
f_1^{BPS}(0)=0,~~~~~~~~~~~~~~~f_1^{BPS}(\infty)=1.\ee
Then the YM BPS monopoles (\ref{YM monopol}) (with their Gribov copies ${ \Phi_i} ^{a(n)}$) display an alike good behaviour (disappearing) at the origin of coordinates and at the spatial infinity ($r\to\infty$). The other thing the behaviour of Higgs  vacuum BPS monopole modes $\Phi_{(n)a}$. These diverge at the origin of coordinates, as it follows from (\ref{sc monopol}) and (\ref{bcf0}). 

\medskip The intersesting and important feature of YM BPS monopoles (\ref{YM monopol}) is that they merge (because of Eq. (\ref{bcf1})) with Wu-Yang monopoles $\Phi^{Wa}_i$ \cite{Wu} \footnote{Remember that Wu-Yang monopoles $\Phi^{Wa}_i$ are solutions to the classical equation of motion \cite{LP1,LP2,Pervush2}
$$D^{ab}_k(\Phi_i^W)F^{bk}_a(\Phi_i^W)=0 \Longrightarrow \frac {d^2f}{d r^2}+\frac {f (f^2-1)}{r^2}=0  $$
in the ``pure'' YM theory (with absent Higgs and fermionic modes) corresponding to the exact $SU(2)$ gauge group. 

One can distinguish  three solutions to this equation: 
$$f_1^{PT}=0,\quad f_1^{W}=\pm 1\quad (r\neq 0). $$
The  first, trivial, solution $f_1^{PT}=0$ corresponds  to the naive unstable perturbation theory, involving  the asymptotic freedom formula \rm
\cite{Mat,Gr}. 
They are just  the Wu-Yang monopoles \cite{Wu} with topological charges $\pm 1$, respectively.
}.

\bigskip The Bogomol'nyi equation (\ref{Bog}) can be treated as a {\it potentiality condition} for the BPS monopole vacuum. A brief  argumentation in favour of this statement was advanced in the recent paper \cite{rem1}. 

Really, mathematically, any potentiality
condition may be written down as
\be \label{potcon} {\rm rot}  ~{\rm grad} ~{ \Phi}=0 \ee
for a scalar field $\Phi$. Thus any  potential field may be represented as ${\rm grad} ~{ \Phi}$.\par
In the Minkowskian YMH theory involving BPS monopole solutions there exists always such  scalar fields. There are just  Higgs  vacuum BPS monopole modes (\ref{sc monopol}) (with their Gribov  topological copies $\Phi_{(n)a}$ \cite{Arsen}).  \par
Then it is easy to guess  that the Bogomol'nyi equation (\ref{Bog}), having the look (\ref{potcon}), can be treated  as the  potentiality condition for the Minkowskian YMH vacuum involving vacuum BPS  monopole solutions. It is so due to the Bianchi identity $D B=0$ \footnote{This becomes more transporent upon representing the Bogomol'nyi equation (\ref{Bog}) in the tensor shape \cite{Al.S.} 
$$ \frac {1}{2g}\epsilon ^{ijk}F_{jk}^a =\nabla ^i\Phi^a.  $$
Then due to the Bianchi identity
$$\epsilon ^{ijk}\nabla _i F_{jk}^b =0,$$
 the   Bogomol'nyi equation (\ref{Bog})  results $$D^2\Phi\sim {\rm rot}{\bf B}=0$$
 (at neglecting the items in $DB$ directly proportional to $g$ and $g^2$).
}.

\medskip Indeed, there can be drawn a highly transparent parallel between the  Minkowskian YMH vacuum involving vacuum BPS  monopole solutions and a liquid helium II specimen described in the Bogolubov-Landau model \cite {N.N.}. \par
In the latter case, the potential motion is proper to the superfluid component in  this liquid helium  specimen. \par
The superfluid motion in a liquid helium II is the motion without  a friction between the superfluid component and the walls of the vessel where a liquid helium  specimen is contained. \par
Thus the viscosity of the superfluid
component in a helium II  is equal to zero, and vortices (involving ${\rm rot}~ {\bf v}\neq 0$) are absent in the superfluid component of a helium. \par
As L. D. Landau   showed \cite {N.N.}, at velocities of the liquid exceeding a {\it critical velocity} $ v_0= {\rm min}~ (\epsilon/p)$ for the ratio of the  energy $\epsilon$ and momentum $p$ for quantum excitations spectrum in
the liquid helium II, the dissipation of the liquid helium energy occurs via arising excitation quanta with momenta $\bf p $ directed antiparallel to the velocity vector $\bf v$. Such dissipation of the liquid helium energy becomes advantageous \cite{Halatnikov} just at
$$ \epsilon+ {\bf p~v}<0 \Longrightarrow \epsilon -p~v<0.$$
From the above reasoning concerning properties of potential motions, it becomes obvious that the vector ${\bf v}_0$ of the critical velocity for the superfluid potential motion possesses the zero curl: ${\rm rot}~ {\bf v}_0=0$. \par
In this case,  according to  (\ref {potcon}), the critical velocity ${\bf v}_0$ of the superfluid potential motion in a liquid helium specimen may be represented \cite{Landau52} as 
\be \label{alternativ} {\bf  v}_0 =\frac{\hbar}{m} \nabla \Phi(t,{\bf r}),
\ee
where $m$ is the mass of a helium atom and $\Phi(t,{\bf r})$
is the phase of the complex-value helium Bose condensate wave function $\Xi (t,{\bf r})\in C$.
\par
Thus the similar look for the vacuum "magnetic" field $\bf B$ in the Minkowskian Higgs model involving   BPS monopole solutions, generating by the Bogomol'nyi equation (\ref{Bog}), and for the critical velocity ${\bf v}_0$ of the superfluid motion in a liquid helium II, given by Eq. (\ref{alternativ}), testifies in favour of the  potential motions occurring therein. \par
In this case, drawing a highly transparent parallel between the  Minkowskian YMH vacuum  involving BPS monopole solutions and a liquid helium II specimen described in the Bogolubov-Landau model \cite {N.N.}, we can also conclude about manifest superfluid properties of the Minkowskian YMH vacuum  involving BPS monopoles. \par
As in the Bogolubov-Landau model \cite {N.N.} 
of  liquid helium II, the ground cause of the superfluid properties of the
Minkowskian YMH vacuum with BPS monopoles roots in  long-range correlations of local excitations \cite {Pervush1}.\par
While in the Bogolubov-Landau model \cite {N.N.} of  liquid helium II this comes to repulsion forces between helium atoms as the cause of superfluidity effects, in the  Minkowskian YMH vacuum involving BPS monopole solutions, the cause of the superfluidity taking place is in the strong YMH coupling $g$ (entering effectively the appropriate action functional (\ref {YM L})). \par
\medskip
The principal thing in alike superfluid effects occurring in a liquid helium II specimen as well as in the Minkowskian YMH vacuum involving BPS monopoles is that these both physical systems are {\it non-ideal gases}. \par
In ideal gases no superfluidity phenomena are possible. \par
There can be demonstrated \cite {Levich} 
that in ideal gases a deal of particles is accumulated on the  zero energy quantum level at temperatures $T<T_0$; herewith the  temperature $T_0$
\be
\label{condensation temperature}
kT_0= \frac{1}{(2.61)^{2/3} }\frac{h^2}{2\pi m} (\frac{N}{V})^{2/3}
\ee
(with $k$ and $h$ being, respectively, the Boltzmann and Planck constants; $N$ being the complete number of particles; $V$  being the volume occupied by the
ideal Bose gas; $m$  being the mass of a particle)
 is called \it the condensation temperature\rm, while the above  deal of particles is called \it the Bose  condensate\rm. \par
 
 \medskip The just described superfluidity is absent in another Minkowskian Higgs models with monopoles: for instance, in the 't Hooft-Polyakov model \cite{H-mon, Polyakov}.  This can be argued, repeating the arguments  \cite{rem1,Hooft}, by disapearing the covariant derivative $D^i\phi_a$ of a Higgs 't Hooft-Polyakov monopole mode $\phi_a$ at the spatial infinity. 
 
 In this case, asymptotically (at $r\to\infty$),
\be \label{3.5}
{\bf B}^a_i D^i\phi_a ={\partial}_i({\bf B}^i_a\phi_a)=0,
\ee
because of the Bianchi identity $DB=0$ and the remark that ${\bf B}^i_a\phi_a$ is a
$U(1)\subset SU(2)$ scalar;  thus one can replace the covariant derivative $D$ with the partial one, $\partial$, for ${\bf B}^i_a\phi_a$. \par
In turn, the complete energy of the YMH configuration may be represented as \cite{Al.S., Hooft}
\be \label{complet}
E_{\rm compl} = \int d^3x~ [\frac{1}{2}(D\phi_a \pm {\bf B}_a)^2+ \frac{\lambda}{4}((\phi^a)^2- a^2)]+ \frac{4\pi}{g^2} M_W.
\ee
The last item in Eq. (\ref{complet}) involves the mass $ M_W $ of the
$W$-boson. \par
Such look of $ E_{\rm compl} $ originates from the paper \cite {H-mon} devoted to the 't Hooft-Polyakov model. \par
The connection between the energy integral $ E_{\rm compl} $ and the general action functional (\ref {YM L}) \cite {LP1,LP2} of the Minkowskian Higgs model is given by the identity \cite{Hooft} 
 \be \label{togd}
(D\phi_a)^2+{\bf  B}_a^2=
(D\phi_a \pm {\bf B}_a)^2 \mp 2 {\bf B}_a D\phi_a.
\ee
Herewith the last item on the right-hand side of (\ref{togd}) vanishes at the spatial infinity, as we have noted above. Just from Eq. (\ref{complet})  one can read  the Bogomol'nyi equation in the shape (\ref{Bog}).

In the 't Hooft-Polyakov model \cite {H-mon, Polyakov} the Bogomol'nyi equation (\ref{Bog}) determines the Bogomol'nyi bound \cite{Hooft} 
\be \label{Emin1}
M_{\rm mon} =\frac{4\pi}{g^2} M_W
\ee
for the complete energy $ E_{\rm compl}$, (\ref{complet}), of the YMH configuration at going over to the BPS limit (\ref{lim}) \cite{BPS}.
\par
Then  the asymptotic  $D_i\phi^a\to 0$ as $ r\to \infty $ for 't Hooft-Polyakov monopoles \cite {H-mon, Polyakov} forces to vanish identically the first item under the integral sign in $ E_{\rm compl}$ ($\vert {\bf B}\vert =0$).\par
In the light of the said above it becomes obvious that the vacuum "magnetic" field $\bf B$, playing the role of the (critical) velocity for the superfluid motion in the Minkowskian non-Abelian vacuum with BPS monopoles, actually approaches zero in the 't Hooft-Polyakov model \cite {H-mon, Polyakov}, involving the $D_i\phi^a\to 0$ as $ r\to \infty $ asymptotic  for Higgs monopoles.  \par

\bigskip The principal goal of the present note is to show that BPS ansatzes $f^{BPS}_{1}(r)$, $f^{BPS}_{0}(r)$, one encounter in the Higgs BPS monopole  model, can serve as electric form-factors therein. Unlike this,  electric form-factors become trivial in the 't Hooft-Polyakov theory \cite {H-mon, Polyakov}.  Grounding this fact will be the topic of Section 2.

In Section 3 we show that presence of BPS ansatzes in the considered  model implies violating the CP invariance. It is the effect similar to that taking place in the instanton models \cite{Cheng,Ryder,Al.S.,Bel}, generating the $\theta$-items in the appropriate effective Lagrangians.  This  effect violating the CP invariance by the $\theta$-dependence of the instanton models was analyzed in the paper \cite{197909065}, and we reconstruct partially the arguments  \cite{197909065} in Section 3.

\section{BPS ansatzes as electric form-factors.}
The starting point of our discussion will be the well known Dirac quantization condition \cite{Dirac} for the electric and magnetic charges presented in a closed system of quantum fields. 

In a simple case when a  quantum object is isolated from another, the Dirac quantization condition \cite{Dirac} acquires the look \cite{Cheng,Ryder,Al.S.} 
\be\label{Dqc}
\frac{q{\bf m}}{4\pi}=\frac 1 2 n; \quad n\in {\bf Z},
\ee
where $q$ and ${\bf m}$ are, respectively, the electric and magnetic charges of the considered object (in the system of units in which $\hbar=c=1$) \footnote{In some sources (for instance, \cite{Ryder,Al.S.}) Eq. (\ref{Dqc}) is given in the slightly modified air 
$$ q {\bf m}= \frac 1 2 n.$$
Going over from Eq. (\ref{Dqc}) to the latter Eq. can be achieved \cite{Hooft} at setting ${\bf m}=2\pi/q$.  The origin of this in the Laplace equation \cite{Ryder}
$$ \nabla \cdot {\bf B}=4\pi{\bf m}\delta^3(r)$$
for the point magnetic charge ${\bf m}$ creating the radial magnetic field ${\bf B}$,
resulting the total magnetic flux 
$$ \Phi= 4\pi r^2B= 4\pi{\bf m}$$
 through a sphere with its centre  in the origin of coordinates.
 
 Just  this provides (see \S 10.3 in \cite{Ryder}) the   change
 $$ \Delta\alpha\vert_\pi = \frac q{\hbar c} \oint {\bf A}\cdot{\bf dl}\vert_\pi=\frac q{\hbar c}\int {\rm rot} {\bf A}\cdot{\bf dS}\vert_\pi=\frac q{\hbar c}\int {\bf B}\cdot{\bf dS}\vert_\pi=\frac q{\hbar c}\Phi(r,\theta)\vert_{\theta=\pi}=\frac q{\hbar c}4\pi{\bf m}=2\pi n$$
 of the dyon's wave function ($\psi\equiv\vert \psi\vert e^{i\alpha}=\vert \psi\vert\exp[(-iq/ \hbar c) {\bf A}\cdot{\bf r}]$) phase $\alpha$ at $\theta=\pi$ (the flux $\Phi(r,\pi)$ is just the maximal possible flux,
   spreeded to the whole sphere). 

}.
Each such quantum object possessing the electric and magnetic charges simultaneously is referred to as a {\it dyon} in modern physical literature \footnote{Formally, a particle possessing the zero electric and magnetic charges also relates to the class of dyons. And moreover, if ${\bf m}=0$, Eq. (\ref{Dqc}) is satisfied at arbitrary values of the electric  charge $q$.}. 

\medskip When a system of quantum fields consists of two dyons, Eq. (\ref{Dqc}) can be generalized to Eq. \cite{Hooft,197909065}  
\be \label{Dir.con}
q_1{\bf m}_2\pm q_2{\bf m}_1 = 2\pi n; \quad n\in {\bf Z};
\ee
Eq. (\ref{Dir.con}) was derived for the first time by Zwanziger and Schwinger \cite{ZwSchw}. The reasoning for deriving this Eq. is \cite{197909065} the classical formula for the angular momentum of an electromagnetic field. The angular momentum in an electromagnetic background of a two-particle system can be calculated easily. It has the magnitude 
$$ (e_1{\bf m}_2- e_2{\bf m}_1)/4\pi c,  $$
that takes integer or half-integer values, as it is expected in quantum mechanics, only  if 
$$ (e_1{\bf m}_2- e_2{\bf m}_1)/\hbar c =2\pi n $$     
(with setting $\hbar= c=1$). Going over to the sign $+$ in (\ref{Dir.con}) from the $-$ one is reduced simply to replacing ${\bf m}\leftrightarrow -{\bf m}$.

In Ref. \cite{Hooft} it was given the in definite  sense generalization of Eq. (\ref{Dir.con}) to the case of an arbitrary gauge group $SU(N)$:
$$ \sum\limits _{i=1}^{N-1} e_i {\bf m}_i=2\pi n.$$
Herewith it is easy to see that  Eq. (\ref{Dir.con}) (with the $+$ sign) is the particular case of the latter relation for the gauge group $SU(2)$.

\bigskip Let us now calculate, following \cite{Cheng}, the total momentum 
\be\label{tmom} {\bf J}={\bf L}+{\bf T}\ee
of a particle in a magnetic monopole background. It involves its spatial angular momentum ${\bf L}$ (including its "ordinary" spin) and the generator $\bf T$ of the internal (for instance, the gauge $U(1)$) symmetry.

On the other hand, ${\bf L}={\bf r}\times{\bf p}$, with $\bf p$ being the canonical momentum
\be\label{Cam} 
p_i= mv_i+g({\bf A}_i\cdot {\bf T})
\ee
for a (vacuum) "gauge" monopole solution involving the mass $m$.

In the particular case of ' t Hooft-Polyakov monopole solutions \cite{H-mon,Polyakov}, when YM potentials have the look \cite{Ryder}
\be 
\label{Polyakov p}
A_i^a= \epsilon _{iab} \frac{r^b}{gr^2}, \ee
Eq. (\ref{Cam}) can be rewritten as 
\be\label{Cam1} 
{\bf p}=m{\dot{\bf r}}+\frac 1{r}  {\bf n}\times{\bf T}.
\ee
Then
$$ 
{\bf J}={\bf r}\times{\bf p}+{\bf T}=
  {\bf r}\times m{\dot{\bf r}}+ {\bf n}\times{\bf n}\times{\bf T}+{\bf T}=$$
\be\label{Jtotal} 
{\bf r}\times m{\dot{\bf r}}+  ({\bf n}\cdot{\bf T}){\bf n}.
\ee
Now we  must recall that in the ' t Hooft-Polyakov model \cite{H-mon,Polyakov}  the radial "magnetic" field $\bf B$ is given by Eq. \cite{Ryder} \footnote{Note that Eqs. (\ref{Polyakov p}) and (\ref{radial pole}) correspond to the "standard" normalization YMH Lagrangian density \cite{Ryder} involving the coefficient $-1/4$ in front of  $F_{\mu \nu}^2$.}
\be  
\label{radial pole}
 F^{ij}= - \frac{1}{gr^3}\epsilon^{ijk} r_k;  \quad {\bf B}_k\equiv -F^{ij}\epsilon_{ijk}.
\ee
From the general reasoning \cite{Cheng} about magnetic monopoles, the equation of motion for an electric  charged particle in its field  is read as
\be\label{mom}
m\ddot{\bf r}=e\dot{\bf r}\times{\bf B}.
\ee
It is just the Lorenz force acting onto this particle in the magnetic monopole background.\par 
In this case the rate of change of the particle angular momentum $\bf L$ is
\be\label{rangu}
\frac d {dt} ({\bf r}\times m\dot{\bf r})={\bf r}\times m\ddot{\bf r}= r^{-3}\frac{2\pi n}{\nu} {\bf r}\times (\dot{\bf r}\times{\bf r})=\frac d {dt} (\frac{2\pi n}{\nu} {\bf n}),
\ee
where \cite{Al.S.} $\nu$ is the (minimal) positive number for which the condition $\exp (\nu h)=1$ (with $h\equiv h(\Phi)\equiv \Phi/a$ being the generator of the residual $U(1)$ gauge group in the quested YMH model) is satisfied \footnote{Defining $\nu$ in this way, one can argue \cite{Al.S.} that 
$$ {\bf m} (\Phi,A)= C~ \zeta (\Phi,A), \quad \zeta (\Phi,A)\in {\bf Z} $$
for the given monopole YMH configuration $(\Phi,A)$ with $C=\nu/4\pi$.}. 

Then, issuing from the above discussed Dirac quantization condition 
\be\label{Dqc1}
q {\bf m}= \frac 1 2 n
\ee
and the normalizations  \cite{Al.S.}
\be \label{eg}
q=\frac{2\pi n}{\nu} g,
\ee
\be \label{mg}
{\bf m}=\frac{\nu}{4\pi g}~\zeta;\quad \zeta \in {\bf Z};
\ee
for the electric and magnetic charges, respectively (thus the  Dirac quantization condition (\ref{Dqc1}) is satisfied automatically), we just arrive to Eq. (\ref{rangu}) (with $q$ given in (\ref{eg}) and appropriate cancelling the YM coupling constant $g$, which enters the relation for $\bf B$).

\medskip The said suggests the formal possibility to introduct the total momentum (\ref{tmom}) \cite{Cheng} of an electric charged
particle in the magnetic monopole background in such a wise that it is conserved:
\be\label{Jtotal1}
{\bf J}={\bf r}\times m{\dot{\bf r}}-\frac{2\pi n}{\nu} {\bf n}= {\bf r}\times m{\dot{\bf r}}-(q/g){\bf n}.
\ee
Comparing then the expressions (\ref{Jtotal}) and (\ref{Jtotal1}) for the total momentum $\bf J$ of an electric  charged particle in the magnetic monopole background got in the ' t Hooft-Polyakov model \cite{H-mon,Polyakov}, one can conclude that
\be\label{posle sravnenija}
(e/g)=-{\bf n}\cdot{\bf T}
\ee
if $q=e$ ($e$ is the elementary charge).

At the particular choice $\bf n$ to be the $z$-direction, ${\bf n}\cdot{\bf T}=T_3$. On the other hand, the value $2\pi/\nu$ can be normalized as $2\pi/\nu=1$ (at considering \cite{Al.S.} the $U(1)$ group space as the circle $S^1$ of the unit radius).

Because of (\ref{eg}), we can conclude  that the isospin operator $\bf T$ ($T_3$) is topologically degenerated. Geometrically, such topological degeneration of the isospin operator $\bf T$ means extracting ("large" and "small") gauge orbits in the $U(1)$ group space. 

Specifying \cite{Cheng} the electric charge operator $Q_{U(1)}=eT_3$, we see additionally that $Q_{U(1)}$ takes integers multipliers of $e$ (at setting $2\pi/\nu=1$)  in the presence of ' t Hooft-Polyakov monopole modes \footnote{This effect was noted, for example, in Ref. \cite{Hooft} with that important correction that besides $Q_{U(1)}=eT_3=ne$ charged states, $({n}+
1/2)e$ states are also possible, with $q=e/2$ being the minimal charge corresponding to  ${\bf T}=1/2$.

For trivial topologies $n=0$, the minimal charge $q=e/2$ corresponds to a fermionic field $\psi$ in the
\begin{eqnarray*}
I=\frac 1 2 : \quad\psi={\psi_1 \choose \psi_2}
\end{eqnarray*}
representation
of $SU(2)$ (if an YMH model is in question).
}.

\medskip
The presence of YM BPS ansatz (\ref{YM monopol}) in the YMH model with (vacuum) BPS monopole solutions changes the computations \cite{Cheng} regarding the isospin operator $\bf T$ and the total momentum $\bf J$.

So instead of (\ref{Jtotal}), then it should be written down
\be\label{JBPStotal}
{\bf J}={\bf r}\times m{\dot{\bf r}}+f_1^{BPS}({\bf n}\cdot{\bf T}){\bf n}+{\bf T} (1-f_1^{BPS}).
\ee
The third item appearing in (\ref{JBPStotal}) corresponds to the expression
$$  {\bf p}=  m{\dot{\bf r}}+(\frac 1{r} f_1^{BPS}) {\bf n}\times{\bf T}  $$
for the momentum $\bf p$ of a particle in the YM BPS monopole background. 

\medskip Indeed, the presence of the YM BPS ansatz (\ref{YM monopol}) complicates  to a considerable extent the computations comparing to those  (\ref{rangu})- (\ref{Jtotal1}) \cite{Cheng} in the ' t Hooft-Polyakov monopole model \cite{H-mon,Polyakov}. It is associated, for instance, with the more  complicated expression for $\bf B$ (see e.g. \cite{Gold}) in the BPS monopole theory. 

But,  in spite of these difficulties, one can conclude, issuing from (\ref{JBPStotal}), that $f_1^{BPS}$ really plays the role of an electric form-factor in the BPS monopole  YMH model. It is because the second item in (\ref{JBPStotal}) can be represented as 
\be \label{form} f_1^{BPS} T_3{\bf n}, \ee
that implies, as it is easy to understand, the replacement $e\leftrightarrow f_1^{BPS}e$ in Eq. (\ref{JBPStotal}). And this is just equivalent to the role of the YM BPS ansatz $f_1^{BPS}$, (\ref{YM monopol}), as an electric form-factor screening the (elementary) charge $e$.

\medskip The  similar role of an electric form-factor is played also by the Higgs BPS ansatz $f_0^{BPS}$, (\ref{sc monopol}).  The physical consequence of  this is   screening effect for electric charges in the Higgs phase \cite{Hooft} of the YMH model, additional to that  rendering in the Higgs phase by the in average electrically neutral Higgs Bose condensate.

A further very important conclusion can be drawn from our discussion in the present section. The presence of BPS ansatzen $f_0^{BPS}$, $f_1^{BPS}$ \cite{BPS}  (in the topologically nontrivial, $n\neq 0$, sectors) of the theories describing the YM-Higgs BPS monopole vacuum turns such theories into specific {\it dyonic} theories since it can be argued (due to Eq. (\ref{form}) above) that some electric charge must be present in the model in discussion. But some warning should be made here. As Eq. (\ref{bcf1}) points, $f_1^{BPS}(0)\to 0$ (and then it is not important that formally electric charges can be present in the model).  This implies the following two cases. In the first case, in the zero topological sector, the spatial distribution for the magnetic charge ${\bf m}({\vec x})$ can take, near the origin of coordinates, any value due to the $0/0$ uncertainty  in which the Dirac quantization condition (\ref{Dqc}) turns in that case. As in the case of Dirac monopoles, investigated by this author in the work \cite{Abel} (see Section 3 ibid), such vacuum configurations involving arbitrary magnetic charges, we will call  {\it magnons}. Vice versa, far from the origin of coordinates (in fact, when $r\to \infty$), in which case $f_1^{BPS}(r)\to 1$,  ${\bf m}({\vec x})\to 0$. This is in a good agreement with the $O(1/r)$ behavior of the average squared of the "magnetic field" $\bf B$:
$$ \int d^3 x (B_i^a)^2\equiv V<B^2>; $$ 
  on the other hand,
$${\bf m}= \frac 1{4\pi} \oint {\bf B} d {\bf S}.$$

One can conclude that   divergences in this "first" case occurs only for  electric charges distributed according to the form-factor 
$f_1^{BPS}({\vec x})$.

\medskip The second case, of nontrivial topologies, is more interesting. At the origin of coordinates, where $f_1^{BPS}(0)\to 0$, now ${\bf m}(0)\to \infty$ according to  (\ref{Dqc}). This implies, at first sight, the UV divergence of the   "magnetic field" $\bf B$ (and thus instability of the whole model) due to the standard definition of the magnetic charge above. But this danger is only apparent indeed.

The fact is that, in the model \cite{LP1,LP2,Pervush2, David2,David3} of the YMH BPS monopole vacuum (in the Dirac quantization scheme \cite{Dirac}) the "magnetic tension" $\bf B$ suffers a discontinuity at the origin of coordinates due to the first order phase transition taking place in the discussed model.

More exactly (see discussion in the papers \cite{disc, sel}), this discontinuity at the origin of coordinates is associated with (topologically nontrivial) thread “counterparts” of YM BPS monopole solutions $\Phi_i^{a{\rm BPS}}$:
\be 
\label{Ateta}
  A  _\theta (\rho, \theta, z)=   \exp(iM\theta) A  _\theta (\rho) \exp(-iM\theta);  \quad  A  _\theta (\rho) = M+ \beta (\rho)\ee 
	with $M$ being the generator of the group $G_1$ of rigid rotations compensating changes in the vacuum YMH  ``thread'' configuration $(\Phi^a,A_\mu^a)$  at rotations around the  axis $z$ of the chosen (rest) reference frame. Here $\Phi^a$  are \cite{Al.S.} z-invariant (vacuum) Higgs solutions in a (small) neighbourhood of the origin of coordinates ($\rho \to 0$):
	\be \label{Higgs-teta}
  \Phi^{(n)}_a (\rho, \theta, z)= \exp (M\theta)~ \phi_a  (\rho) \quad (n\in {\bf Z}), \quad \nabla_\mu \phi(\rho) \leq {\rm const}~\rho ^{-1-\delta}; \quad  \delta>0; \quad n\in {\bf Z};\ee
 $\rho=\sqrt{x^2+y^2} $ is the distance from the axis $z$.  The function $\beta (\rho)$ in (\ref{Ateta}) approaches zero as $\rho \to \infty$. \par

It is obvious that rectilinear threads $A_\theta $ don't coincide with vacuum YM BPS monopole solutions $\Phi_i^{a{\rm BPS}}$ \cite{LP1,LP2} (thus they don't belong to the "{\it BPS spectrum}"  involving  the electric form-factor $f_1^{BPS}$), and, on the contrary, there are gaps between directions of "magnetic" tensions vectors: ${\bf B}_1$,
\be \label{B1} \vert {\bf B}_1 \vert \sim \partial _\rho A_\theta (\rho,\theta, z), \ee 
and $\bf B$, given by the Bogomol'nyi equation (\ref{Bog}) (and diverging as $r^{-2}$ at the origin of coordinates).

In the  model \cite{LP1,LP2,Pervush2, David2,David3} for the YMH BPS monopole vacuum, the vacuum expectation value (VEV) $<B^2>$ plays the role of the order parameter of that model, and the above gap just testifies in favor of the first order phase transition taking place therein. On the other hand, in such way one can avoid the above mentioned infrared instability (for nontrivial topologies) of the discussed model.

Indeed, this is equivalent to the statement that
 \be \label{B1v}
<B_1^2> =0,
\ee
i.e. $$<\partial_ \rho A_\theta (\rho,\theta, z)> =0\Leftrightarrow <\partial_\rho \beta(\rho)>=0\Leftrightarrow < \beta(\rho)>={\rm const}.$$
Thus the first order phase transition condition in the discussed Minkowskian YMH model  can be reduced to the simple condition of (nonzero) VEV for $\beta(\rho)$ given in (\ref{Ateta}).

On the other hand, as it was argued in the papers \cite{disc,sel}, the alternative physical sense of the latter condition is that it determines  the {\it false vacuum} in the Minkowskian YMH model quantized by Dirac (at the  simultaneous interpretation of the VEV of the "magnetic field" ${\bf B}_1$ squared as the order parameter for the rotary phase inside that vacuum). Meanwhile the {\it true} vacuum of the model \cite{LP1,LP2,Pervush2, David2,David3}  is determined by the value of $<B^2>\ne 0$ set by the  Bogomol'nyi equation (\ref{Bog}). This vacuum possesses superfluidity-like properties as it was argued in \cite{rem1}. This just implies the first order phase transition taking place in the model \cite{LP1,LP2,Pervush2, David2,David3} for the YMH BPS monopole vacuum.

Due to the standard definition of the magnetic charge $\bf m$ above, one can assert that Eq. (\ref{B1v}) is equivalent to the assertion  

\be \label{m1}
<{\bf m}_1>=0,
\ee
where $<{\bf m}_1>$ is the vacuum average value of the magnetic charge (distribution)  ${\bf m}_1({\vec x})$ specifying the "magnetic field" ${\bf B}_1$.  Then the Dirac quantization condition 
(\ref{Dqc1}) implies that  the "rotary component" of the discussed YMH vacuum (involving the "thread" configuration  $(\Phi^a,A_\mu^a)$ \cite{Al.S.}) is is in fact (in averaging) a purely electric medium for the trivial topology $n=0$. At $n\neq 0$ (it is again a purely electric case), in turn, the model acquires (infrared) divergences and requires a renormalization in this case. 

Note that the condition (\ref {m1}) promotes the asymptotic freedom of quarks in the spatial region near the origin of coordinates. It  promotes the asymptotic freedom of quarks along with the effect annihilating topologically nontrivial YM modes of the same topological number (say, $n$) at colliding with (topologically nontrivial) threads $A_\theta$ \cite{Al.S., disc}.
\section{Higgs BPS ansatzes and CP violating.} 
In this section, repeating the arguments \cite{197909065}, we shall attempt to demonstrate that the presence of YM BPS ansatz $f_1^{BPS}$, (\ref{YM monopol}), in the (Minkowskian) YMH BPS monopole theory violates manifestly the CP invariance of that theory. 

 \medskip In the previous section we have discussed the Dirac-Zwanziger-Schwinger quantization condition (\ref{Dir.con}) \cite{197909065}. It turns out that this condition says something important about the difference between electric
charges of two  magnetic monopoles.

 Given, for example, two monopoles of minimum allowed magnetic charge $2\pi/e$ and of electric charges $q$ and $q^{'}$, one finds 
 \be
 \label{3.1}
 e_1{\bf m}_2- e_2{\bf m}_1= 2\pi(q-q^{'})/e,
 \ee
 so that the Dirac-Zwanziger-Schwinger quantization condition (\ref{Dir.con}) gives
 \be
 \label{3.2}
  q-q^{'}=ne.
 \ee
 Thus the difference $q-q^{'}$ must be an integer multiple of  $e$. But as it was noted in  \cite{ZwSchw}, there is no restriction onto $q$ and $q^{'}$ separately. 

If, however, the Dirac-Zwanziger-Schwinger quantization condition (\ref{Dir.con}) is suplemented by the CP conservation, the allowed values of the electric charge of an magnetic monopole are also quantized. 

In fact, although the electric charge is odd under CP, the magnetic charge is even (this is because  electric and magnetic fields are transformed  oppositely under parity). Applied to a monopole of the  charges $(q,2\pi/e)$, a CP transformation gives the monopole of the  charges $(-q,2\pi/e)$. For these two particles 
\be
 \label{3.3}
 e_1{\bf m}_2- e_2{\bf m}_1=4\pi q/e
 \ee
 and is a multiple of $2\pi$ only if
 \be
 \label{3.4}
 q=ne\quad {\rm or} \quad q=(n+1/2)e.
 \ee
 Thus at assuming the CP conservation, monopoles can have integer or half-integer  electric charges (as multiples of  $e$).  And moreover, if monopoles of  integer electric charges exist, monopoles of  half-integer charges do not exist and vice-versa.
 
 \bigskip  Appart from the CP conservation, there are no reasoning for satisfying the claim (\ref{3.4}). In nature the CP invariance is violated, but weakly. One can thus suspect  that monopoles possess almost (half)integer electric charges. The deviation of monopoles from such  charges would be proportional to the strenght of CP violating \cite{197909065}. 
 
\medskip The one  source  CP violating is the instanton YM model \cite{Bel}, resulting  \cite{Cheng} the effective Lagrangian 
\be \label{eL}
{\cal L}_{\rm eff} \equiv {\cal L}+\Delta {\cal L}={\cal L}+\frac{g^2\theta}{16\pi^2} ~{\rm tr}~ ( F_{\mu \nu}^a \tilde F^{\mu \nu});  \quad \tilde F_{\mu \nu}= \frac 1 2 \epsilon^{\mu \nu \alpha\beta} F_{\alpha\beta};
\ee 
involving the quasimomentum $\theta\in [-\pi, \pi]$ \footnote{Indeed, as it was argued in the papers \cite{ Arsen,Pervush1,Galperin} (see also \cite{rem3}), $\theta$ is a complex parameter.

This can be seen at performing the quantization procedure for the instanton YM model \cite{Bel}. The latter one is reduced to solving the system of equations \cite{ Arsen,Pervush1,Galperin,rem3} 
\be
{\hat H}(i\delta/\delta A, A) \Psi _\epsilon[A]=\epsilon \Psi _\epsilon[A], \nonumber \ee 
\be 
\nabla _i^{ab}(A) (\frac {\delta}{i\delta A_{i b}}\Psi_\epsilon [A])=0; \quad \nabla _i^{ab}(A)= \delta ^{ab}\partial_i- g\epsilon ^{abc}A_{ic}; \nonumber
\ee    
\be 
T_1 \Psi _\epsilon [A]= e^{i 1\cdot \theta} \Psi_\epsilon[A].\nonumber
\ee 
for the wave function $\Psi _\epsilon [A]$, the quantum analogue of an instanton  \cite{Bel} possessing the energy $\epsilon$.

The first equation in this system is the Schr$\rm\ddot{o}$dinger equation for the YM Hamiltonian 
$$ {\hat H}= \int  d^3x \frac {1}{2} [E^2+B^2]; \quad E= \frac {\delta}{i\delta A}.$$
The second one expresses  the normalization assumed in the instanton YM model \cite{Bel} at which the electric field $E$ is transverse \cite{Arsen}:
$$ \nabla _i E^i\Psi _\epsilon [A]=0. $$
At last, the third equation implicates the {\it raising operator} \cite{Cheng} 
$$  T_1 \vert n>=  \vert n+1>$$
(with the winding number 1 as its eigenvalue). In the terminology  \cite{Fadd2}, the transverse electric field $E$ remains invariant with respect to all the ``large'' transformations  $T_1$, while the instanton wavefunction $\Psi _\epsilon [A]$ is manifestly covariant with respect to these transformations.

The raising operator $T_1$ can be represented explicitly in the shape \cite{Pervush1} 
$$  T_1= \exp (\frac {d}{dX[A]})= \exp \left\{\left[\int d^3x B^2 \frac {g^2}{16\pi^2}\right]^{-1} \int d^3x B_i^a \frac {\delta}{\delta A_i^a}\right\},$$
with $X[A]$ being the YM winding number functional (its look is well known and we will not cite it here).

\medskip The above system of equations gives a correct definition (cf. \cite {Callan1}) of the $\theta$-vacuum as a pseudomomentum operator possessing the common system of eigenfunctions $\{\Psi _\epsilon [A]\}$ with the momentum operator $\nabla _i^{ab}(A)$.

Indeed, one encounters the following problem with this definition \cite{Pervush1,Callan1} of the $\theta$-vacuum. The thing is that the operators $\hat H$ and $T_1$ don't commute (as it was argued in \cite{Pervush1}) unless $\epsilon=0$, i.e. the have no common eigenfunction $\Psi _\epsilon [A]$ at $\epsilon\neq 0$ (on the contrary, the Hamilton operator $\hat H$ commutes with the momentum $\nabla _i^{ab}(A)$). 

It is obvious \cite{Pervush1, Galperin,rem3} that at setting $\epsilon=0$, the definition \cite{Pervush1,Callan1} of the $\theta$-vacuum  remains valid for imaginary as well as for real values of $\theta$. This allows to represent $\theta$ as a complex number \cite{rem3} $\theta=\theta_1+i\theta_2$.
In particular, at purely imaginary values of $\theta=\theta_2$ and $\epsilon=0$,  there exists the plane wave 
$$ \Psi _0[A]= \exp \{i(2\pi k +\theta_2)X[A]\}\equiv\exp \{iP_{\cal N}X[A]\}  \quad (k\in {\bf Z})  $$
which satisfies formally our definition \cite{Pervush1,Callan1} of the $\theta$-vacuum at 
$$ P_{\cal N}=2\pi k +\theta_2=2\pi k\pm i 8\pi^2/ g^2.  $$
Here the real part of $P_{\cal N}$, $\theta_1$, runs formally over the discrete set $2\pi k$, while its imaginary part $\theta_2=8\pi^2/ g^2$ is  continuous. 

But this plane wave functional diverges manifestly at the $-$ sign before $8i\pi^2/ g^2$  (this fact was pointed out, for instance, in the papers \cite{LP2,rem3}). 

Simultaneously, one can constract (repeating the arguments \cite{Pervush1,Galperin}; see also \cite{rem3}) the family of purely real solutions for the topological momentum $P_{\cal N}$ satisfying the Schr$\rm\ddot{o}$dinger equation at $\epsilon=0$  being parallel (via the $\theta$) the eigenvalue of the raising operator $T_1$: 
$$  P_{\cal N}^{\rm R}=2\pi k \pm 8\pi^2/ g^2. $$
Issuing from this Eq., one can  equate $\theta_1=\pm 8\pi^2/ g^2$ and assume $\theta_1$ to vary in the interval $[-\pi,\pi]$; it is just that {\it real} $\theta$-angle considered in modern gauge physics (see, for instance, \cite{Cheng}).

On the other hand, the presence of imaginary (i.e. space-like) momentum modes $8i\pi^2/ g^2$ in the $ P_{\cal N}$ spectrum (with the appropriate ``diverged'' plane waves) gives it impossibly to give the correct probability description of the (topologically degenerated) $\theta$-vacuum sinse the Hilbert space $\Psi _0^{(n)} [A]$ of its states becomes non-separable in this case.

\medskip The remarkable property of the real part $P_{\cal N}$ spectrum is following \cite{Pervush1,rem3}. It is obvious that such real topological momentum $P_{\cal N}^{\rm R}$ vanishes (i.e. the instanton YM configuration stops) in the limit $g\to\infty$ for the YM coupling constant $g$. It is just the infrared QCD confinement limit as it is understood in modern physic.  In the terminology \cite{Pervush1} this case is referred to as the  {\it infrared catastrophe}. 

Note that purely imaginary (and thus space-like and unphysical) values $P_{\cal N}=\pm  8\pi^2 i/ g^2$ (at $k=0$) have no relation to the infrared catastrophe.

\bigskip The results we have demonstrated now can be  treated  \cite{Pervush1} as the presence of unphysical solutions to the  Schr\"odinger equation at the application of ordinary quantization methods to a topologically nontrivial theory. In the paper \cite{Arsen}  this statement was referred to as the so-called \it no-go theorem\rm. \par 
}. 

The  name ``quasimomentum'' for $\theta$ has the following origin. The  thing is that even real values  of the topological momentum $P_{\cal N}$ have rather a fictive nature. This is so since the $\theta$-term in the instanton YM effective Lagrangian (\ref{eL}) does not alter  \cite{Al.S.} the YM  equations of motions 
$$   D_\mu F^{\mu\nu}=0.     $$
 It is obvious that the  degree one of the quasimomentum $\theta$ in the instanton YM effective Lagrangian (\ref{eL}) implies its manifest P (and CP because of this) covariance. 

 The question about the ``separate'' C-covariance of the $\theta$-item in (\ref{eL}) is more delicate.  To understand which a ``delicacy'' is concealed here let us now resort to the arguments  \cite{197909065}. 

In this paper the effect influence of the (real) $\theta$-angle upon the dyon charge was investigated. 

To determine the concrete effect of the CP violation of the dyon charge by $\theta$-angle (latter proves to be conserved if $\theta=0$; we have made sure in this above), one must apply a semi-clasical analysis. For instance, the author  \cite{197909065}  utilize the simple semi-clasical analysis has been performed in the work \cite{Tomb}.

In this work a semi-clasical quantization of clasical dyonic solutions was performed. In a gauge in which fields disappear at the (spatial) infinity,  clasical dyonic solutions are periodic in   time. The semi-clasical quantization condition comes to the claim \cite{197909065,Tomb} that $S+ET$ (the action  during the time period $T$ plus the energy times the  time) should be an integer multiple of $2\pi$. 

Let $I$ being   the action per unit time. Then the above claim ``that $S+ET$ should be an integer multiple of $2\pi$'' is reduced to the relation 
$$ T(I+E)=2\pi n \quad (n\in{\bf Z}).$$
The clasical period $T$ and the ``abbreviated action'' \cite{197909065} $I+E$ (at $T=1$) were calculated in Ref. \cite{Tomb} in the absence of the CP violation. It was found that 
\be \label{3.6}
I+E=cq^2;
\ee 
\be \label{3.7}
T=\frac{2\pi}{ec}~~\frac 1 q,
\ee
where $q$ is the electric charge of the dyon an $c$ is a constant \footnote{It is not a large problem to calculate this constant, but a simple arguments shows that the same constant $c$ appears in Eqs. (\ref{3.6}) and (\ref{3.7}).}.

The condition $T(I+E)=2\pi n$ now gives 
\be \label{3.8}
q=ne,
\ee
so that dyons possess integer electric charges (in the $e$ units) as one expects in the absence of the CP violation.

\bigskip Let us assume now that $\theta\neq 0$ and let us repeat the above calculations. At $\theta\neq 0$ the equations of motion are unchanged \cite{Al.S.,rem3}, and  there no change in the period $T$ or the energy $E$. However, there is an extra contribution to the  action $I$ from the extra item $\Delta{\cal L}$ in the effective instanton YM Lagrangian (\ref{eL}) \footnote{In the light of our above conjecture \cite{Pervush1,Galperin,rem3} that $\theta=\theta_1+i\theta_2$, the expression below, recast to the look
$$ E= cq^2-I+ceq \frac\theta{2\pi}= cq^2-I+ceq \frac{\theta_1+i\theta_2}{2\pi},$$
 shows transparently that the ``complex'' $\theta$-vacuum is not stationar. Due to the ordinary quantum mechanic canons, the life time of a $\theta$-vacuum state with the energy $E$ (indeed, we should set $E=0$ sinse only such $\theta$-vacuum states are compatible with the definition, us given above, of this vacuum) is 
 $$ \tau=\frac{2\pi\hbar}{\theta_2} $$
 (appropriately, its line width is $\theta_2/2\pi$). 

This means that the $\theta$-vacuum state with the energy $E=0$ decays into (two) states: say, 1 and 2, for which $E_1+E_2=0$. But these $\theta$-vacuum states, involving nonzero energies $E_1$ and  $E_2$, are badly  specified. As we have emphasized  above, repeating the arguments  \cite{Pervush1},  these values of energy cannot be simultaneously common eigenvalues of the YM Hamilton operator $\hat H$ and the raising operator $T_1$.

In this is the  next in turn contradiction about the YM instanton model. And moreover, the  life time $\tau$ of a $\theta$-vacuum state with the energy $E=0$ is a finite number as $\theta_2\neq 0$.  Thus the $E=0\to E_1+E_2$ vacuum decay occurs in the finite time $\tau$. This implies once again bad specifying the quantum states $\vert 1>$ and $\vert 2>$ corresponding to the energies $E_1$ and  $E_2$ as those not referring to the time infinities.  

As it was discussed in Ref. \cite{rem3} (repeating  the said in the monograph \cite{Logunov}), in that case $\tau\neq\pm\infty$ it is impssible to describe correctly  Feynman diagrams  referring to the above $E=0\to E_1+E_2$ vacuum decay since only at the claim $\tau\to\pm\infty$ the interaction representation of the system of quantum fields, set with the aid of the appropriate scattering matrix $S$, is true. If  $\tau\neq\pm\infty$, one cannot pick out (as a consequence of the Haag theorem \cite{Logunov}) a correct Fock representation for  interacting (quantum) fields.  

 }:
 \be \label{extra}
 I+E=cq^2+ceq \frac\theta{2\pi}. 
 \ee
The semi-classical (Bohr-Sommerfeld) quantizaation of $T(I+E)$ then gives 
\be \label{3.10}
q=ne-\theta e /2\pi.
\ee
So the allowed values of magnetic monopoles (if latter ones are contemplated in the model: for instance if Higgs modes are present in such model, violating  the initial $SU(2)$ gauge symmetry)  are not integer if $\theta\neq 0$. In particular, there does not exist an electrically neutral magnetic monopole in that case (since now $q$ never equal to zero). 

The zero topology case $n=0$ becomes an especial one in this context in the instanton YM model involving the $\theta$ angle. 

Let us reconsider again the Dirac quantization condition (\ref{Dqc}). At  $n=0$ and $q\neq 0$ (it is just the case at $\theta \neq 0$), the magnetic charge $\bf m$ will be also zero \footnote{If $q=0$
 and $n=0$, the magnetic charge $\bf m$ can take any value due to the $0/0$ uncertainty. Such purely magnetic configurations can be referred as "magnons". Also it is interesting to note that for the system of two dyons at $n=0$, appropriate Eq. (\ref{Dir.con}) can result zero at $q_1\neq 0$ and  $q_2\neq 0$ (as just in the  $\theta\neq 0$ situation discussed above) in the following cases. If $q_1=q_2$, it is possible  when ${\bf m}_1=\pm{\bf m}_2$ as it can be seen from 
(\ref{Dir.con}). Also, it is possible for the general case $q_1\neq q_2$ if ${\bf m}_1={\bf m}_2=0$.}. 

If no magnetic and electric charges are presented in the gauge theory (i.e. there are no magnetic monopoles in that theory): for instance in the ``pure'' YM theory without Higgs modes (the instanton  model is just such a case), one would set $q=0$ in (\ref{3.10}) (this does not contradict the Dirac quantization condition (\ref{Dqc}) since one deals with the uncertainty $0/ 0$ in that case). Then 
\be \label{3.12}
n=[\theta/2\pi],
\ee
i.e. the integer part of the number $\theta/2\pi$. It is, in fact, the ordinary connection between the $\theta$-angle and the integer topological number $n$ in the instanton  model.

\bigskip  Now we should recall that in the absence of magnetic monopoles the $\theta$-angle results no ``true'' motions in the YM theories. Instead  of this, the $\theta$-dependense in a YM model comes to purely tunnelling effects connected with instantons \cite{Cheng,Bel} (for the topological number $n=1$ such effects of the order  $\exp(1/\alpha)$; $\alpha\equiv g^2/4\pi$).

But if magnetic monopoles (for instance, of the 't Hooft-Polyakov type \cite{H-mon,Polyakov}) are incorporated in the YM theory, in the monopole sector of that theory there are classically allowed motions, dyons, with the nonzero topological number
$$ n\sim\int d^4x F_{\mu \nu}\tilde F^{\mu \nu} \in {\bf Z}.$$
As a result, the $\theta$-dependense in the monopole sector is something another than that connected with instantons: in particular, it is of the leading order rather different  from $\exp(1/\alpha)$.

  \bigskip What happens, asks the author \cite{197909065}, in  the YM theories in which CP is violated by  another mechanism than $\theta$? 
  
  The fact that at $\theta=0$ the dyons have integer charges is associated with the fact that $I+E$ in Eq. (\ref{3.6}) is quadratic in $q$, with no linear term. A  linear term as in Eq. (\ref{extra}) leads to noninteger charges: this is directly proportional to the coefficient of the linear term. 
  
  CP forbids such linear term sinse $q$ is odd under CP. If CP is violated, regardless the violation mechanism, a linear term can be present. 
  
  Even if a linear term is absent on the classical level (for instance, if only couplings to fermions violate CP, since fermions do not enter classical  solutions), it can be present on the quantum level due to loop corrections. Roughly speaking, one should recalculate $I+E$ from the {\it quantum} effective action  (rather than from the {\it classical} action).  If CP is violated, loop corrections to the effective action would induce a term linear in $q$ in the  effective $I+E$ and therefore to cause monopole charges (if exist) to be not quite integer. 
  
  \medskip The same effect CP violating occurs, as we have demonstrated in the previous section, in the YMH model involving BPS magnetic monopoles (without any $\theta$-dependense). The crucial point here is the presence of the BPS ansatz $f_1^{BPS}$, (\ref{YM monopol}), in that model (as it can be seen from (\ref{JBPStotal})).
  But these BPS magnetic monopole solutions induce (as it can be demonstrated) rather tree than loop Feynman diagrams.
  \section{Discussion.} The Dirac fundamental quantization \cite{Dir} of the YMH model involving vacuum BPS  monopole solutions (coming, as we have explained above, to solving the Gauss law constraint in terms of topological Dirac variables) implies some refining us said in the previous section about the  CP violation  in the presence of those vacuum BPS  monopole solutions.
  
  As it was discussed  in the recent paper \cite{rem3}, resolving the YM Gauss law constraint
  
  \be \label{Gauss} \frac {\delta W}{\delta A^a_0}=0 \Longleftrightarrow [D^2(A)]^{ac}A_{0c}= D^{ac}_i(A)\partial_0 A_{c}^i. \ee
 in terms of topological Dirac variables 
  \be
 \label{s12}
 \hat A_i^{ (n)}(t,{\bf x}) = \hat\Phi_i^{ (n)}({\bf x}) + {\hat{\bar A}}_i^{ (n)} (t,{\bf x}); \quad n\in {\bf Z}; \quad\hat A_\mu = g \frac {A_\mu^a\tau _a}{2i \hbar c}
\ee
(here ${\hat{\bar A}}_i^{ (n)} (t,{\bf x})$ are perturbation excitations, {\it multipoles}, over the  BPS  monopole vacuum); satisfying the Coulomb gauge $D^i { \hat A_i} ^{a(n)}=0$, results the homogeneous equation
 \be \label{Gaussh}
 [D^2(A)]^{ac}\hat A_{0c}=0
 \ee
 for temporal components $\hat A_{0c}$ of YM fields. 
 
 As it was shown in Ref. \cite{Pervush1}, the solutions to the ``homogeneous'' YM Gauss law constraint (\ref{Gaussh}) can be found in the shape 
  \be \label{zero}  A_0^c(t,{\bf x})= {\dot N}(t) \Phi_{(0)}^c ({\bf x})\equiv Z^c,
   \ee
implicating the topological dynamical variable $\dot N(t)$ and Higgs (topologically trivial) vacuum Higgs BPS monopole modes $\Phi_0^a ({\bf x})$.

Issuing from {\it zero mode solutions} \cite{Pervush1} $Z^a$ to the YM Gauss law constraint (\ref{Gauss}), it is easy to  write down $F_{i0}^a$ components of the YM tension tensor, taking the shape of so-called vacuum "electric" monopoles \cite{LP1,LP2} 
$$   F^a_{i0}={\dot N}(t)D ^{ac}_i(\Phi_k ^{(0)})\Phi_{0c}({\bf x}).         $$
In turn, vacuum "electric" monopoles $F^a_{i0}$ generate the action functional
\be \label{rot} W_N=\int d^4x \frac {1}{2}(F_{0i}^c)^2 =\int dt\frac {{\dot N}^2 I}{2}\ee
involving the  {\it rotary momentum} \cite{David2}
\be \label{I} I=\int \sb {V} d^3x (D_i^{ac}(\Phi_k^0)\Phi_{0c})^2 =
\frac {4\pi^2\epsilon}{ \alpha _s}
=\frac {4\pi^2}{\alpha _s^2}\frac {1}{ V<B^2>}.    \ee
This action functional (describing the solid rotations of the YMH vacuum suffered the Dirac fundamental quantization) is manifestly Poincare (P) invariant sinse it can be recast to the look 
\be
 \label{rotat}
 W_{N}= \int dt  \frac{ P_N^2 (t)}{2I}; \quad P_N={\dot N}I.
 \ee
 Herewith 
\be \label{pin} P_N ={\dot N} I= 2\pi k +\theta; \quad \theta  \in [-\pi,\pi];      \ee 
is the momentum corresponding to the abovementioned solid rotations of the YMH vacuum. As it can be seen from (\ref{pin}), the $P_N$ spectrum is purely real and thus rotary trajectories for the YMH vacuum suffered the Dirac fundamental quantization belong to the physical ones. 

It is just the momentum spectrum free from imaginary, i.e. tachionic, modes,  unlike the instanton case, us discussed in Section 3. 

 \medskip Besides that the action functional (\ref{rotat}) is manifestly P-invariant, it is also C-invariant. To understand this fact, we should compare Eq. (\ref{rotat}) (it is a definite functional of the YMH BPS monopole vacuum rotary energy $P_N^2 (t)/2I$) with Eq. (\ref{extra}) \cite{197909065}. 
  
In the latter case the appropriate energy $I+E$ associated with the instanton $\theta$ vacuum proves to be linear (i.e. odd) by $e$ and $\theta$.  This just implies CP violating.

The former case is rather different. 
As it follows from  (\ref{I}), (\ref{rotat}), the action functional (\ref{rotat}) is directly proportional to the vacuum "electric" field ("electric" monopole) $F^a_{i0}$, induced, in turn, by a Higgs vacuum  BPS monopole mode $\Phi_{0c}({\bf x})$. 

As it was shown in Ref. \cite{David2} (see also \cite{rem3}), 
\be
\label{se} F^a_{i0}\equiv E_i^a=\dot N(t) ~(D_i (\Phi_k^{(0)})~ \Phi_{(0)})^a= P_N \frac {\alpha_s}{4\pi^2\epsilon} B_i^a (\Phi _{(0)})= (2\pi k +\theta) \frac {\alpha_s}{4\pi^2\epsilon} B_i^a(\Phi_{(0)}).
       \ee
       Latter relations were got with the aid of the Bogomol'nyi equation (\ref{Bog}). Sinse the YM coupling constant $\alpha_s\equiv g^2/4\pi(\hbar c)^2$ enters Eq. (\ref{se}), it is already squared by $g$, while the rotary momentum $I$, (\ref{I}), is of the order $g^4$. 
       
       But at accepting the normalization (\ref{eg}) \cite{Al.S.} for the electric charges $q$ (referring to the Higgs Bose condensate), the  vacuum "electric" field   $F^a_{i0}$ will be of the order $q^{-2}$ (respectively, the YMH BPS monopole vacuum rotary energy $P_N^2 (t)/2I$ will be of the order $q^{-4}$).  Thus the YMH model with vacuum  BPS monopoles quantized by Dirac is C-invariant as that possessing  the action functional (\ref{rotat}) even by the total electric charge $Q=\sum_i q_i $ of (topologically degenerated) Higgs vacuum  BPS monopole modes $\Phi_{c(n)}({\bf x})$ \cite{Arsen}. 
       
 This us demonstrated CP conservation is the next in turn remarkable feature of that  model.      

\end{document}